\documentclass[1p]{elsarticle}
\usepackage{amsthm}  % Load amsthm first
\usepackage{amsmath,amssymb,amsfonts,graphicx,bbm,mathtools,braket,color}

\newtheorem{theorem}{Theorem}[section]
\newcommand{\gf}[1]{{#1}}
\newcommand{\gff}[1]{{#1}}
\newcommand{\mga}[1]{{#1}}
\begin{document}
\begin{frontmatter}
\title{Cartan Quantum Metrology}
\author{Gabriele Fazio$^1$}
\ead{gabriele.fazio@studenti.unimi.it}
\author{Jiayu He$^2$}
\ead{jiayu.he@helsinki.fi}
\author{Matteo G. A. Paris$^1$}
\ead{matteo.paris@fisica.unimi.it}
\address{${}^1$Dipartimento di Fisica {\em Aldo Pontremoli}, Universit\`a 
degli Studi di Milano, I-20133 Milano, Italy}
\address{${}^2$QTF Centre of Excellence, Department of Physics, 
University of Helsinki, FI-00014 Helsinki, Finland}
%%%
\begin{abstract}
We address the characterization of two-qubit gates, focusing on bounds 
to precision in the joint estimation of the three parameters that 
define their Cartan decomposition. We derive the optimal probe states 
that jointly maximize precision, minimize sloppiness, and eliminate 
quantum incompatibility. Additionally, we analyze the properties of the 
set of optimal probes and evaluate their robustness against noise.
\end{abstract}
\date{\today}
\end{frontmatter}
%%%%%%%%%%%%%%%%%%%%%%%%%%%%%%%%%%%%%%%%%%
\section{Introduction}
The characterization of two-qubit gates is critical to any protocol in 
quantum information processing \cite{Kraus2001,Vidal2004,Vatan2004}. 
Two-qubit gates form the fundamental 
building blocks for universal quantum computation, enabling more 
complex operations through entanglement between qubits 
\cite{Barenco1995,Makhlin2002,Zhang2003,rezakhani2004characterization}. Precise 
characterization of two-qubit gates enables the design of accurate gates, as even small errors can propagate and degrade the overall performance of quantum 
algorithms \cite{Gilchrist2005}. 
Furthermore, accurate gate characterization is essential 
for error correction protocols \cite{Bennett1996} and, in turn, 
for building fault-tolerant quantum computers \cite{Chow2012} 
and scalable quantum systems. 

The Cartan decomposition represents a powerful 
tool for analyzing generic two-qubit gates by breaking them down 
into simpler, more fundamental components \cite{Khaneja2001,shende2004minimal,Tucci05,PhysRevA.87.012106}. 
According to Cartan decomposition, 
any two-qubit gate can be expressed as a product of local single-qubit 
operations and a non-local part, known as the Cartan's kernel. The 
kernel contains the essential entangling operations %\gff{,}
and is characterized 
by three independent parameters. Since local single-qubit operations 
can be efficiently controlled and do not contribute to entanglement, 
only the Cartan kernel is truly relevant for studying the gate’s 
features.  In other words, the Cartan kernel captures 
the essential properties of 
the gate and simplifies the problem to the estimation of three 
parameters, which govern the non-local behavior of the gate.

In this paper, we address the joint estimation of the three Cartan 
parameters of a two-qubit gate \cite{Liu_2019,Albarelli_2020} 
by preparing an initial probe state, 
allowing it to interact with the gate, and then perform %ing
measurements 
on the resulting output state. Our goal is to optimize the encoding 
process so that all parameters are estimable, minimizing the sloppiness 
of the model while maximizing precision. Additionally, we aim to avoid 
any extra quantum noise that may arise from the non-commutativity of the measurements \cite{Razavian_2020}. As we will demonstrate, 
it is possible to jointly achieve 
these three objectives, setting new benchmarks for the precise 
characterization of two-qubit quantum gates and providing 
insights into the precision-sloppiness tradeoff in qubit systems.

The paper is structured as follows. In Section \ref{s:not} we introduce 
notation and the Cartan’s decomposition theorem applied 
to elements of $SU(4)$, whereas in Section \ref{s:qet} we briefly 
reviews the tools of multiparameter quantum metrology. 
In Section \ref{s:cartanm}, we find the optimal states to achieve 
minimum sloppiness at fixed precision, whereas in Section \ref{s:noise} 
we assess the robustness of optimal probes in the presence of 
noise. Section \ref{s:outro} closes the paper with some 
concluding remarks. 

%%%%%%%%%%%%%%%%%%%%%%%%%%%%%%%%%%%%%%%%%%
\section{Cartan decomposition}\label{s:not}
A generic state of a %qbit
\gf{qubit} system may be written in the Bloch representation 
as 
\begin{equation}
\rho = \frac12 \left( %\mathbb{I}
\gf{\mathbb{I}_2} + \underline{r} \cdot 
\underline{\sigma} \right)\,,
\end{equation}
where the Bloch vector $\underline{r}$ is given by
\begin{equation}
	\underline{r} = \left(\text{Tr}[\sigma_x \rho], 
		\text{Tr}[\sigma_y \rho],\text{Tr}[\sigma_z \rho]\right)^T
		\,,
\end{equation}
and $\underline{\sigma}=(\sigma_x, \sigma_y, \sigma_z)^T$ is the vector of Pauli matrices. 
The purity of the state $\rho$ is linked to the Bloch vector as
\begin{equation}
	\mu[\rho] = \frac{1}{2}(1+|\underline{r}|^2)\,.
\end{equation}

Given $n$ %qbits
\gf{qubits}, the state of the whole system lives in the tensor product of the 
Hilbert spaces $\mathcal{H}^n = \bigotimes_{k=1}^n \mathcal{H}_k$.
\gf{In the specific case of two qubits \gff{prepared} in a pure state, we can use the concurrence as a measure of entanglement\cite{Wootters2001}. Given a generic state} %the concurrence is given by}
$\ket{\psi} = \sum_{x,y \in \{0,1\}} \alpha_{x,y}\ket{x}\ket{y}$ %to be
% \gf{it's defined as}
\gff{, the concurrence is defined as}
\begin{equation}
	\mathcal{C}(\ket{\psi}) \equiv 2 |\alpha_{00}\alpha_{11}-\alpha_{01}\alpha_{10}|.
\end{equation}
This can be generalized for mixed states by
%$\mathcal{C}(\rho) \equiv \text{max}(0,\lambda_1-\lambda_2-\lambda_3-\lambda_4)$, where
%$\lambda_1\geq\lambda_2\geq\lambda_3\geq\lambda_4$ are the eigenvalues of operator
\gff{$\mathcal{C}(\rho) \equiv \text{max}(0,\Lambda_1-\Lambda_2-\Lambda_3-\Lambda_4)$}, where
\gff{$\Lambda_1\geq\Lambda_2\geq\Lambda_3\geq\Lambda_4$} are the eigenvalues of operator
$R \equiv \sqrt{\sqrt{\rho}(\sigma_y\otimes\sigma_y)\rho^*(\sigma_y\otimes\sigma_y)\sqrt{\rho}}$.

\begin{figure}[h!]
	\centering\includegraphics[width=0.8\textwidth]{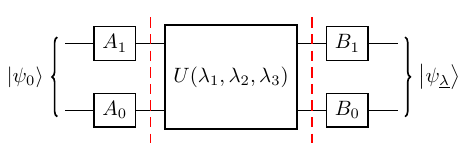}
	\caption{The decomposition of a generic gate into
	single \gf{qubit} operations and a Cartan kernel.}
	\label{fig:CartanDecomposition}
\end{figure}

Let us now consider a generic unitary gate acting on two %qbits
\gf{qubits}. Such an 
evolution corresponds to a linear operator in $SU(4)$. Élie Joseph Cartan 
proved the following theorem.
\begin{theorem}[Cartan's KAK decomposition]
	Given a group $\underline{G} = \text{exp}(\underline{g})$, with a subgroup
	$\underline{K} = \text{exp}(\underline{k})$, and a Cartan subalgebra
	$\underline{a}$, where $\underline{g} = \underline{k} \oplus \underline{k}^\perp$
	and $\underline{a} \subset \underline{k}^\perp$.\\
	Then any $G \in \underline{G}$ can be written as $G = K_1 A K_2$,
	with $K_1,K_2 \in \underline{K}$ and $A \in \text{exp}(\underline{a})$.
\end{theorem}
A special case of this theorem has been developed by Khaneja and Glaser
for quantum computing. See \cite{Tucci05} for a 
constructive proof using only linear algebra, of what is referred \gff{to} as KAK1 theorem.
\begin{theorem}[Khaneja and Glaser's KAK1 theorem\cite{Khaneja2001}]
	Given any $S \in SU(4)$, there exists $A_0,A_1,B_0,B_1 \in SU(2)$,
	and $\underline{k} \in \mathbb{R}^3$, such that
	\begin{equation}
		S = (A_1 \otimes A_0)e^{i\underline{k}\cdot\underline{\Sigma}}(B_1\otimes B_0)
	\end{equation}
	where $\Sigma \equiv \left(\sigma_x\otimes\sigma_x,
		\sigma_y\otimes\sigma_y,\sigma_z\otimes\sigma_z\right)^T 
        \equiv
		\left(\sigma_{XX},\sigma_{YY},\sigma_{ZZ}\right)^T$.
	\label{th:KAK1}
\end{theorem}
Starting from a generic operator over $\mathbb{C}^4$, which has 16
free components, unitarity reduces the degrees of freedom only by 1.
One of the strengths of this theorem is the ability to separate the
contribute of all these different parameters. In fact\gff{,} what we see is
that only 3 of the initial 15 parameters are involved in the interaction
between the two %qbits
\gf{qubits}. The other 12 corresponds to single %qbit
\gf{qubit} evolutions.
This decomposition is depicted in %Figure
\gf{Fig.} \ref{fig:CartanDecomposition}.

The $U$ gate is referred to as the \gf{Cartan's kernel},
and will be the main object of our analysis.
Using theorem \ref{th:KAK1} we may write the gate $U$ as 
\begin{equation}
	U 
	= \exp\left\{
	 - i \sum\limits_j \lambda _j \sigma_j\otimes\sigma_j \right\}\,,
	 \label{gate}
\end{equation}
which, in the standard basis, reads as 
follows (where $\lambda_\pm = \lambda_1 \pm \lambda_2$)
\begin{equation}
U= \left(
		\begin{array}{cccc}
			e^{-i \lambda _3} \cos \lambda_- & 0 & 0 & -i e^{-i \lambda _3} \sin \lambda_-  \\
			0 & e^{i \lambda _3} \cos \lambda_+ & -i e^{i \lambda _3} \sin \lambda_+& 0 \\
			0 & -i e^{i \lambda _3} \sin \lambda_+ & e^{i \lambda _3} \cos \lambda_+ & 0 \\
			-i e^{-i \lambda _3} \sin \lambda_- & 0 & 0 & e^{-i \lambda _3} \cos \lambda_- \\
		\end{array}
		\right)\,.
\label{eq:U}
\end{equation}
 
Using instead the Bell basis
%\begin{equation}
%\ket{\beta_{jk}} = \Big(\frac{\ket{0}\ket{k}+(-1)^j \ket{1}\ket{\bar{y}}}{\sqrt{2}}\Big)_{C}\,,
%\gf{\ket{\beta_{kj}} = \Big(\frac{\ket{0}\ket{k}+(-1)^j \ket{1}\ket{\bar{k}}}{\sqrt{2}}\Big)_{C}\,,}
$\gf{\frac1{\sqrt{2}}\left\{
        \ket{00}+\ket{11},
       \ket{00}-\ket{11},
     \ket{01}+\ket{10},
 \ket{01}-\ket{10}
        \right\}}
$, 
%\end{equation}
$U$ has the following diagonal form

\begin{equation}
%	U = \hbox{Diag}\left(
%		 e^{-i (\lambda_1-\lambda_2+\lambda_3)},
%		e^{-i (\lambda_1+\lambda_2-\lambda_3)} ,
%		 		e^{i (\lambda_1-\lambda_2-\lambda_3)}, 
%		e^{i (\lambda_1+\lambda_2+\lambda_3)} \right)
    \gf{U = \hbox{Diag} 	
	\left\{
		 e^{-i (\lambda_1-\lambda_2+\lambda_3)} ,  e^{i (\lambda_1-\lambda_2-\lambda_3)} , e^{-i (\lambda_1+\lambda_2-\lambda_3)}, e^{i (\lambda_1+\lambda_2+\lambda_3)}		\right\}.}
\end{equation}
The last thing to be discussed about the decomposition is the domain of
the parameters.
We initially have that $\lambda_j \in [0,2\pi) \in \mathbb{R}$.
Then, we can define an equivalence relation, where
two kernels $U,V \in SU(4)$ are said to be
{equivalent up to local operations}, if there exists
$R_0,R_1,L_0,L_1 \in U(2)$ such that $U=(R_1\otimes R_0)V(L_1\otimes L_0)$.
It can be proven that this is a valid equivalence relation.
Starting from this we can find three operations on $\underline{\lambda}$ 
that preserve
the equivalence class. 1) Shift: $\underline{\lambda}$ can be shifted on any of the
	three components by an integer multiple of $\frac{\pi}{2}$. So given
	$n,m,l \in \mathbb{Z}$.
	%\begin{equation}
		%$(\lambda_x,\lambda_y,\lambda_z) \mapsto (\lambda_x+n\frac{\pi}{2},\lambda_y+m\frac{\pi}{2},\lambda_z+l\frac{\pi}{2});$
    %\end{equation}
\gff{$(\lambda_1,\lambda_2,\lambda_3) \mapsto (\lambda_1+n\frac{\pi}{2},\lambda_2+m\frac{\pi}{2},\lambda_3+l\frac{\pi}{2});$}
2) Reverse: any two components can change sign. For example
	%\begin{equation}
		%$(\lambda_x,\lambda_y,\lambda_z) \mapsto (-\lambda_x,-\lambda_y,\lambda_z)$;
	%\end{equation}
\gff{$ (\lambda_1,\lambda_2,\lambda_3) \mapsto (-\lambda_1,-\lambda_2,\lambda_3)$;}
3) Swap: any two components can be swapped. For example
	%\begin{equation}
		%$(\lambda_x,\lambda_y,\lambda_z) \mapsto (\lambda_y,\lambda_x,\lambda_z)$.
	%\end{equation}
\gff{$(\lambda_1,\lambda_2,\lambda_3) \mapsto (\lambda_2,\lambda_1,\lambda_3)$.}
Using these operation it is possible to reduce the domain to what is called
the canonical class vector set
%\begin{equation}
%    \gf{D_C} =\{\underline{\lambda} \in \mathbb{R}^3 \ : \
%	\frac{\pi}{2}>\lambda_x\geq\lambda_y\geq\lambda_z\geq0
%	\wedge \lambda_x+\lambda_y\leq\frac{\pi}{2}\}
%\end{equation}
\begin{equation}
	\gff{
    \gf{D_C} =\{\underline{\lambda} \in \mathbb{R}^3 \ : \
	\frac{\pi}{2}>\lambda_1\geq\lambda_2\geq\lambda_3\geq0
	\wedge \lambda_1+\lambda_2\leq\frac{\pi}{2}\}}
\end{equation}
For more details, see \cite{Tucci05}. 

%%%
\section{Multiparameter quantum estimation}\label{s:qet}
%%%
The estimation of a single parameter or more than one have 
common features, but differ for fundamental reasons 
\cite{brida2010experimental,paris2009quantum}. The first 
one is that in quantum mechanics there may be observables 
which are %non compatible
\gff{incompatible}, e.g., the observables employed to 
estimate  two different parameters \cite{Albarelli_2020}. When this is the case, the 
joint estimation of the  
parameters is unavoidably affected by additional noise 
due to the non-commutativity of the measured observables. Additionally, 
there may be correlations between the parameters, or one of the parameter 
may be more relevant %of
\gf{than} the others, for the problem at hand. In all 
those cases,  a weight matrix should be introduced to optimize precision 
under the given constraints. 
\par
In classical statistics, given a set of parameters to be estimated
$\underline{\lambda}$, from the measurement of one or more 
observables $\underline{X}$, we have a conditional probability
$p(\underline{x}|\underline{\lambda})$. The Fisher information 
matrix is defined by
\begin{equation}
	F_{jk} = \sum\limits_{\underline{x}} \frac{\partial_j p(\underline{x}|\underline{\lambda}) \ \partial_k p(\underline{x}|\underline{\lambda})}{p(\underline{x}|\underline{\lambda})},
	\label{eq:classicalF}
\end{equation}
where $\underline{x}$ denote the outcome(s) of the measurement of 
$\underline{X}$. We denote by $\underline{\hat{\lambda}}\equiv\underline{\hat{\lambda}}(\{x_j\})$, 
an estimator that maps the set of outcomes into
the space of parameters. We also assume to work with unbiased estimators, 
meaning $\braket{\underline{\hat{\lambda}}}=\underline{\lambda}$.
The covariance matrix of the estimator is given by
$\text{Cov}[\underline{\hat{\lambda}}]_{jk} \equiv \braket{\hat{\lambda}_j\hat{\lambda}_k}-
\braket{\hat{\lambda}_j}\braket{\hat{\lambda}_k}$ and the multiparameter classical Cramér-Rao bound \cite{Cramer1999} is the matrix inequality
$\text{Cov}[\underline{\lambda}] \geq \frac1M F^{-1}$, where $M$ denotes the number of repeated measurements.

In order to build the analogue quantum Cramér-Rao bound, we first notice that for a quantum system, the conditional probability is given by the 
Born rule which, for pure 
\gf{states}, reads
$p(\underline{x}|\underline{\lambda}) = |\braket{\underline{x}|\psi_{\underline{\lambda}}}|^2$. We then introduce the symmetric 
logarithmic derivative(SLD) of a parameter, which is implicitly defined by
the relation \cite{Amari2000}
\begin{equation}
	\partial_{\lambda_j} \rho \equiv \partial_j \rho \equiv
	\frac{L_{\lambda_j}\rho+\rho L_{\lambda_j}}{2}
	\equiv \frac{L_j\rho+\rho L_j}{2},
	\label{eq:SLD}
\end{equation}
where the subscript $j$ corresponds to the derivative with respect
to the j-th parameter $\lambda_j$.
In the case of pure states, the solution may be easily obtained by 
observing that $\rho^2=\rho$.  We thus get
\begin{equation}
	\partial_j\rho = \partial_j(\rho^2) =
	(\partial_j\rho) \rho + \rho (\partial_j\rho)\,,
	\label{eq:pureRho}
\end{equation}
and comparing %Equations \ref{eq:SLD} to \ref{eq:pureRho}
\gf{Eq.} (\ref{eq:SLD}) to (\ref{eq:pureRho}) we see that
 a solution
is given by
\begin{equation}
	L_j = 2\partial_j \rho.
	\label{eq:pureSLD}
\end{equation}

The quantum Fisher information matrix(%QFI
\gf{QFIM}) $[Q]_{jk}$ is then defined 
via the SLD by
\begin{equation}
	Q_{jk} \equiv \text{Tr}\Big[\rho \frac{L_j L_k + L_k L_j}{2}\Big]
	=\frac{1}{2}\text{Tr}[\rho\{L_j,L_k\}]\,
	\label{eq:QFI}
\end{equation}
where $\rho$ is the state of the system. In general, it may be difficult to 
obtain the SLDs in operatorial form, and it is often convenient to
express the %QFI
\gf{QFIM} in terms of the eigenvalues and eigenvectors of
the statistical operator, as follows \cite{Liu_2019}
\begin{theorem}
	Given the spectral decomposition of a density matrix
	$\rho = \sum_{y_j \in S} y_j \ket{y_j}\bra{y_j}$,
	where $S = \{y_j \in \{y_j\} |y_j \neq 0\}$
	is the support of non null eigenvalues. Then $Q_{jk}$ is given by
    \gf{
	\begin{align}
   			Q_{jk} = & \sum\limits_{y_l \in S}\Big(
				\frac{(\partial_j y_l)(\partial_k y_l)}{y_l}
				+ 4 y_l \hbox{\rm Re} 
				[\braket{\partial_j y_l|\partial_k y_l}] \Big) \notag \\
				&-\sum\limits_{y_l,y_m \in S}
				\frac{8 y_l y_m}{y_l+y_m}\text{\rm Re} [\braket{\partial_j y_l|y_m}
				\braket{y_m|\partial_k y_l}].
		\label{eq:generalQFI}
	\end{align} }
\end{theorem}
A useful expression for the \gf{QFIM} may be obtained 
for pure states(i.e. $\rho_{\underline{\lambda}}=\ket{\psi}\bra{\psi}$), 
as follows
\begin{equation}
	Q_{jk} = 4\,\hbox{Re} [\braket{\partial_j\psi|\partial_k\psi}-
	\braket{\partial_j\psi|\psi}\braket{\psi|\partial_k\psi}].
	\label{eq:pureQFI}
\end{equation}
The %QFI
\gf{QFIM} is an important object because it enters the quantum version 
of Cram\`er-Rao bound, posing a lower bound on the covariance matrix of the estimators.

The Uhlmann curvature, also called
incompatibility matrix is defined via
the SLD, but instead of using the anticommutator as in %Equation
\gf{Eq.} (\ref{eq:QFI}), the commutator is used.
\begin{equation}
	D_{jk} \equiv -i\text{Tr}\Big[\rho\frac{L_j L_k-L_k L_j}{2}\Big]
	= -\frac{i}{2}\text{Tr}[\rho[L_j,L_k]]
	\label{eq:uhlmann}
\end{equation}
From the definition we see that the main diagonal %is
\gf{elements are} always
made of zeros. The meaning of the off diagonal terms is related
to the quantum incompatibility. For $D=0$, all parameters are 
compatible, and the model is said
to be asymptotically classical \cite{Albarelli_2020,Razavian_2020}. 
In this case, the optimal
estimation is asymptotically achievable via collective measurements. 
For pure states, we may write 
\begin{equation}
	D_{jk} = 4\,\hbox{Im} [\braket{\partial_j\psi|\partial_k\psi}-
	\braket{\partial_j\psi|\psi}\braket{\psi|\partial_k\psi}].
	\label{eq:pureD}
\end{equation}

As in the classical case, we denote by $\underline{\hat{\lambda}}\equiv\underline{\hat{\lambda}}(\{x_j\})$, 
an unbiased estimator that maps the set of outcomes into
the space of parameters. The multiparameter Cramér-Rao bound states that
\begin{equation}
	\text{Cov}[\underline{\lambda}] \geq \frac{Q^{-1}}{M}\,.
\end{equation}
A scalar bound, limiting the overall precision of the multiparameter 
estimation strategy may be obtained \gff{by} taking the trace of the above inequality, thus obtaining a bound on the sum of the variances of 
the estimators. By doing so we have
\begin{equation}\label{sym}
	\text{Tr}\left[\text{Cov}[\underline{\lambda}]\right] 
	%\equiv
    \gf{=} \sum_j \hbox{Var}\,\hat\lambda_j \geq \frac{\text{Tr}[Q^{-1}]}{M}
	\,.
\end{equation}
The quantity $p=\text{Tr}[Q^{-1}]$ provides a lower bound to the optimal overall precision. \mga{As mentioned above, a weight matrix $W$ may be introduced in the definition of the precision $p=\text{Tr}[W\,Q^{-1}]$
to adjust the Cramér-Rao bound for specific applications. However, in our case, we set $W = {\mathbb I}$ (the identity matrix) to emphasize that the three parameters of the Cartan kernel should be treated on the same foot since each one of them provide essential information on the nature of the two-qubit gate at hand.}

In general, the bound in Eq. (\ref{sym}) 
is not achievable and a different scalar bound, termed Holevo bound \cite{HOLEVO1977251}  has been derived, which is achievable asymptotically. In our case, however, the two bounds coincide and therefore the quantity $p$ quantifies the overall precision 
of our scheme (see below for details). 

Another feature of multi-parameter statistical models is that the encoding 
state may be not efficient, meaning that different parameters 
values are assigned locally to the same state and it is impossible to 
accurately recover them from the statistics of data.  In these cases, 
the (classical or quantum) statistical model is termed 
{\em sloppy} \cite{brown2003statistical,brown2004statistical,PhysRevLett.97.150601,machta2013parameter,yang2023untwining,frigerio2024overcoming}
and the (quantum) Fisher Information matrix is singular. 
This means that one or more parameters have zero (quantum) FI, and the 
statistical model is effectively only dependent on the other parameters. 
This may be seen explicitly by a reparametrization, not necessarily a 
linear one. The sloppiness of a statistical model may be quantified 
by the inverse of the determinant of \gf{the} (quantum) Fisher Information matrix
$s= 1/\hbox{Det}\, Q$, and a relevant question in the analysis of a given 
model is whether there is a trade-off between precision and sloppiness 
or \gf{whether} they %maybe
\gf{can be} jointly optimized, i.e., there exist conditions in which 
the trace of the inverse %QFI
\gf{QFIM} is minimum and the determinant of the %QFI
\gf{QFIM}
is maximum. 

To summarize, in our multiparameter \gf{model}, $d=\hbox{Det} D$ 
quantifies the non commuting nature of the different SLDs and, in turn, the presence of additional noise of quantum origin. On the other hand, $p=\text{Tr}[Q^{-1}]$ quantifies the precision achievable in the joint estimation of the parameters, whereas $s= 1/\hbox{Det}[Q]$ quantifies the sloppiness of the model, i.e. the degree of efficiency in the encoding of the parameter on the probe state. In the next Section, we will prove that in our case there are probe 
states leading to $d=0$, and thus no additional quantum noise is expected 
in the joint estimation of the Cartan parameter. Our goal will be therefore 
that of optimizing over the possible probe states in order to jointly 
minimize $p$ (i.e., maximize the precision) and $s$ (i.e., minimize 
sloppiness).  
Notice that the symmetric nature of the %QFI
\gf{QFIM} %\gff{matrix}
$Q$ already poses a bound on precision in terms of sloppiness. In fact,  for a $n \times n$ real symmetric matrix $Q$, we have that $\hbox{Tr} [Q^{-1}] \geq n\, \hbox{Det}[Q]^{-1/n}$. In our $3 \times 3$ case, 
this implies that
\begin{equation}
\label{matb}
 p \geq 3\, s^{1/3}\,.
\end{equation}
We anticipate that the optimal probes found in the next Sections will allow us to saturate this bound.

\section{Cartan metrology}
\label{s:cartanm}
In this Section, we address the estimation \gf{of} the three parameters 
$\underline{\lambda}$ that characterize a Cartan gate of the form (\ref{gate}).
In particular, we seek for the optimal achievable precision 
and for the probe states that allow one to achieve such precision. Additionally, we prove that incompatibility vanishes and analyze the tradeoff between precision and sloppiness of the statistical model. 
\mga{The quantum Fisher information is a convex function of the quantum state, which implies that the maximum QFI is achieved by pure states. Thus, restricting the analysis to pure states identifies the tightest possible Cramér-Rao bound. Due to the extended convexity of the Fisher information \cite{ECSahar}, the same is true for a two-qubit probe. Mixing two-qubit states cannot enhance the total QFI beyond the pure-state limit, and we thus
assume that the two-\gf{qubit} probe is prepared in a pure state.}
%%%
\subsection{Quantum Fisher information matrix \gf{and Uhlmann Curvature}}
To evaluate the %QFI
\gf{QFIM} %\gff{matrix},
we need the form of the state after the gate
$\ket{\psi_{\underline{\lambda}}} \equiv U\ket{\psi_0}$ and, in turn, a parametrization 
of the initial state. In the canonical computational base\gff{,} we have
\begin{equation}
    \ket{\psi_0} \equiv \left(
         \alpha ,
         \beta  e^{i \phi _{\beta }} ,
         \gamma  e^{i \phi _{\gamma }} ,
         \delta  e^{i \phi _{\delta }} 
        \right)^T\,,
\end{equation}
with
$\alpha,\beta,\gamma,\delta \in [0,1]$, and
$\phi_\beta,\phi_\gamma,\phi_\delta \in [0,2\pi)$
such that $\alpha^2+\beta^2+\gamma^2+\delta^2 = 1$.
From now on greek letters will correspond to the parametrization in the canonical base.
Given the expression of the gate transformation $U$ in Eq. (\ref{eq:U}), we get
\begin{equation}
    \ket{\psi_{\underline{\lambda}}} = \left(
        \begin{array}{c}
         e^{-i \lambda_3} \left[\alpha  \cos (\lambda_1-\lambda_2)-i \delta  e^{i \phi _{\delta }} \sin (\lambda_1-\lambda_2)\right] \\
         e^{i \lambda_3} \left[\beta  e^{i \phi _{\beta }} \cos (\lambda_1+\lambda_2)-i \gamma  e^{i \phi _{\gamma }} \sin (\lambda_1+\lambda_2)\right] \\
         e^{i \lambda_3} \left[\gamma  e^{i \phi _{\gamma }} \cos (\lambda_1+\lambda_2)-i \beta  e^{i \phi _{\beta }} \sin (\lambda_1+\lambda_2)\right] \\
         e^{-i \lambda_3} \left[\delta  e^{i \phi _{\delta }} \cos (\lambda_1-\lambda_2)-i \alpha  \sin (\lambda_1-\lambda_2)\right] \\
        \end{array}
        \right).
    \label{eq:finState}
\end{equation}
Using Eq. (\ref{eq:pureQFI}) we obtain the elements of the %QFI
\gf{QFIM}
\gf{
\begin{equation}
    \begin{split}
        %&Q_{11} = 4 \Big[\alpha ^2 \left[1-2 \delta ^2\right]-2 \alpha  \delta  \left[\alpha  \delta  \cos \left(2 \phi _{\delta }\right)+4 \beta  \gamma  \cos \left(\phi _{\delta }\right) \cos \left(\phi _{\beta }-\phi _{\gamma }\right)\right]\\
        %&\ \ \ \ \ \ \ \ \ \ -2 \beta ^2 \gamma ^2 \cos \left(2 (\phi _{\beta }- \phi _{\gamma })\right)-2 \beta ^2 \gamma ^2+\beta ^2+\gamma ^2+\delta ^2\Big]\\
        %&Q_{11}=4 \Big[-2 \alpha ^2 \delta ^2-2 \alpha  \delta  \left[\alpha  \delta  \cos \left(2 \phi _{\delta }\right)+4 \beta  \gamma  \cos \left(\phi _{\delta
   %}\right) \cos \left(\phi _{\beta }-\phi _{\gamma }\right)\right]\\
   %&\ \ \ \ \ \ \ \ \ \ -2 \beta ^2 \gamma ^2 \cos \left(2 (\phi _{\beta }- \phi _{\gamma })\right)-2 \beta ^2
   %\gamma ^2+1\Big]\\
        &Q_{11} = 4 \Big[1-2 \alpha  \delta  \cos \left(\phi _{\delta }\right)-2 \beta  \gamma  \cos \left(\phi _{\beta }-\phi _{\gamma }\right)\Big]\Big[1+2 \alpha  \delta  \cos \left(\phi _{\delta }\right)+2 \beta 
   \gamma  \cos \left(\phi _{\beta }-\phi _{\gamma }\right)\Big]\\
        %&Q_{22} = 4 \Big[\alpha ^2 \left[1-2 \delta ^2\right]-2 \alpha  \delta  \left[\alpha  \delta  \cos \left(2 \phi _{\delta }\right)-4 \beta  \gamma  \cos \left(\phi _{\delta }\right) \cos \left(\phi _{\beta }-\phi _{\gamma }\right)\right]\\
        %&\ \ \ \ \ \ \ \ \ \ -2 \beta ^2 \gamma ^2 \cos \left(2 (\phi _{\beta }- \phi _{\gamma })\right)-2 \beta ^2 \gamma ^2+\beta ^2+\gamma ^2+\delta ^2\Big]\\
        %&Q_{22} = 4 \Big[-2 \alpha ^2 \delta ^2-2 \alpha  \delta  \left[\alpha  \delta  \cos \left(2 \phi _{\delta }\right)-4 \beta  \gamma  \cos \left(\phi _{\delta }\right) \cos \left(\phi _{\beta }-\phi _{\gamma
   %}\right)\right]\\
   %&\ \ \ \ \ \ \ \ \ \ -2 \beta ^2 \gamma ^2 \cos \left(2 (\phi _{\beta }- \phi _{\gamma })\right)-2 \beta ^2 \gamma ^2+1\Big]\\
        &Q_{22} = 4 \Big[1-2 \alpha  \delta  \cos \left(\phi _{\delta }\right)+2 \beta  \gamma  \cos \left(\phi _{\beta }-\phi _{\gamma }\right)\Big] \Big[1+2 \alpha  \delta  \cos \left(\phi _{\delta }\right)-2 \beta 
   \gamma  \cos \left(\phi _{\beta }-\phi _{\gamma }\right)\Big]\\
        %&Q_{33} = 4 \Big[-\left[(\alpha -1) \alpha -\beta ^2-\gamma ^2+\delta ^2\right] \left[\alpha ^2+\alpha -\beta ^2-\gamma ^2+\delta ^2\right]+\beta ^2+\gamma ^2+\delta ^2\Big]\\
        &Q_{33} = 16 \left[\alpha^2 + \delta^2 \right] \left[\beta ^2+\gamma ^2\right]\\
        %&Q_{12} = Q_{21} = 4 \Big[2 \alpha ^2 \delta ^2 \cos \left(2 \phi _{\delta }\right)+\left(2 \alpha ^2-1\right) \delta ^2-\alpha ^2-2 \beta ^2 \gamma ^2 \cos \left(2 (\phi _{\beta }- \phi _{\gamma })\right)\\
        %&\ \ \ \ \ \ \ \ \ \ \ \ \ \ \ \ \ \ \ \ -2 \beta ^2 \gamma ^2+\beta ^2+\gamma ^2\Big]\\
        &Q_{12} = Q_{21} = 4 \Big[-\alpha^2+\beta^2+\gamma^2-\delta^2+4 \alpha ^2 \delta ^2 \cos ^2\left(\phi _{\delta }\right)-4 \beta ^2 \gamma ^2 \cos ^2\left(\phi _{\beta }-\phi _{\gamma }\right)\Big]\\
        %&Q_{13} = Q_{31} = 8 \Big[\beta  \gamma  \left[-\alpha ^2+\beta ^2+\gamma ^2-\delta ^2-1\right] \cos \left(\phi _{\beta }-\phi _{\gamma }\right)\\
        %&\ \ \ \ \ \ \ \ \ \ \ \ \ \ \ \ \ \ \ \ +\alpha  \delta  \cos \left(\phi _{\delta }\right) \left[-\alpha ^2+\beta ^2+\gamma ^2-\delta ^2+1\right]\Big]\\
        &Q_{13} = Q_{31} = 16 \Big[\alpha  \delta  \left(\beta ^2+\gamma ^2\right) \cos \left(\phi _{\delta }\right)-\beta  \gamma  \left(\alpha ^2+\delta ^2\right) \cos \left(\phi _{\beta }-\phi _{\gamma }\right)\Big]\\
        %&Q_{23} = Q_{32} = 8 \Big[\beta  \gamma  \left[-\alpha ^2+\beta ^2+\gamma ^2-\delta ^2-1\right] \cos \left(\phi _{\beta }-\phi _{\gamma }\right)\\
        %&\ \ \ \ \ \ \ \ \ \ \ \ \ \ \ \ \ \ \ \ +\alpha  \delta  \cos \left(\phi _{\delta }\right) \left[\alpha ^2-\beta ^2-\gamma ^2+\delta ^2-1\right]\Big]\\
        &Q_{23} = Q_{32} = -16 \Big[\alpha  \delta  \left(\beta ^2+\gamma ^2\right) \cos \left(\phi _{\delta }\right)+\beta  \gamma  \left(\alpha ^2+\delta ^2\right) \cos \left(\phi _{\beta }-\phi _{\gamma }\right)\Big]
    \end{split}
\end{equation}
}
Despite being seemingly complex, we get symmetrical and
relatively simple expressions for its determinant, and
the trace of its inverse
\begin{align}
    p &= \frac{3}{16} \left(\frac{\alpha ^2+\delta ^2}{\alpha ^4-2 \alpha ^2 \delta ^2 \cos 2 \phi _{\delta }+\delta ^4}+\frac{\beta ^2+\gamma ^2}{\beta ^4-2 \beta ^2 \gamma ^2 \cos \left(2 (\phi _{\beta }- \phi _{\gamma })\right)+\gamma ^4}\right),
    \label{eq:TrIF}
\\
    \frac1s &= 1024 \left[\alpha ^4+\delta ^4-2 \alpha ^2 \delta ^2 \cos 2 \phi _{\delta }\right] \left[\beta ^4+\gamma ^4-2 \beta ^2 \gamma ^2 \cos \left(2 (\phi _{\beta }- \phi _{\gamma })\right)\right].
    \label{eq:DetF}
\end{align}
\mga{
Notice that  the trace of the inverse and the determinant of the %QFI
\gf{QFIM} are not functions of $\underline{\lambda}$, namely the statistical model is {\em covariant}. While this would have been a trivial result if the three generators were commuting operators, this is not the case in our problem.} There are a few implications stemming from this fact. The first is that 
all the $\underline{\lambda}$'s can be estimated with the same precision.
The same observation implies that this is true also for  $\underline{\lambda}=\underline{0}$, i.e. for $U=\mathbb{I}$, showing that optimality is 
achievable independently of the fact that gate has or not an entangling power. 

% \subsection{Uhlmann curvature}
Using Eq. (\ref{eq:uhlmann}) we can calculate the Uhlmann curvature $D$, 
and by doing so we find that $D=0$. As mentioned above, this implies 
that our model is asymptotically classical, meaning that the CRB is 
asymptotically saturable. Additionally\gff{,} this also tells us that the three 
parameters are compatible, and there is no additional noise due to 
quantum incompatibility.

\subsection{State optimization}
\begin{figure}[h!]
	\centering\includegraphics[width=0.75\textwidth]{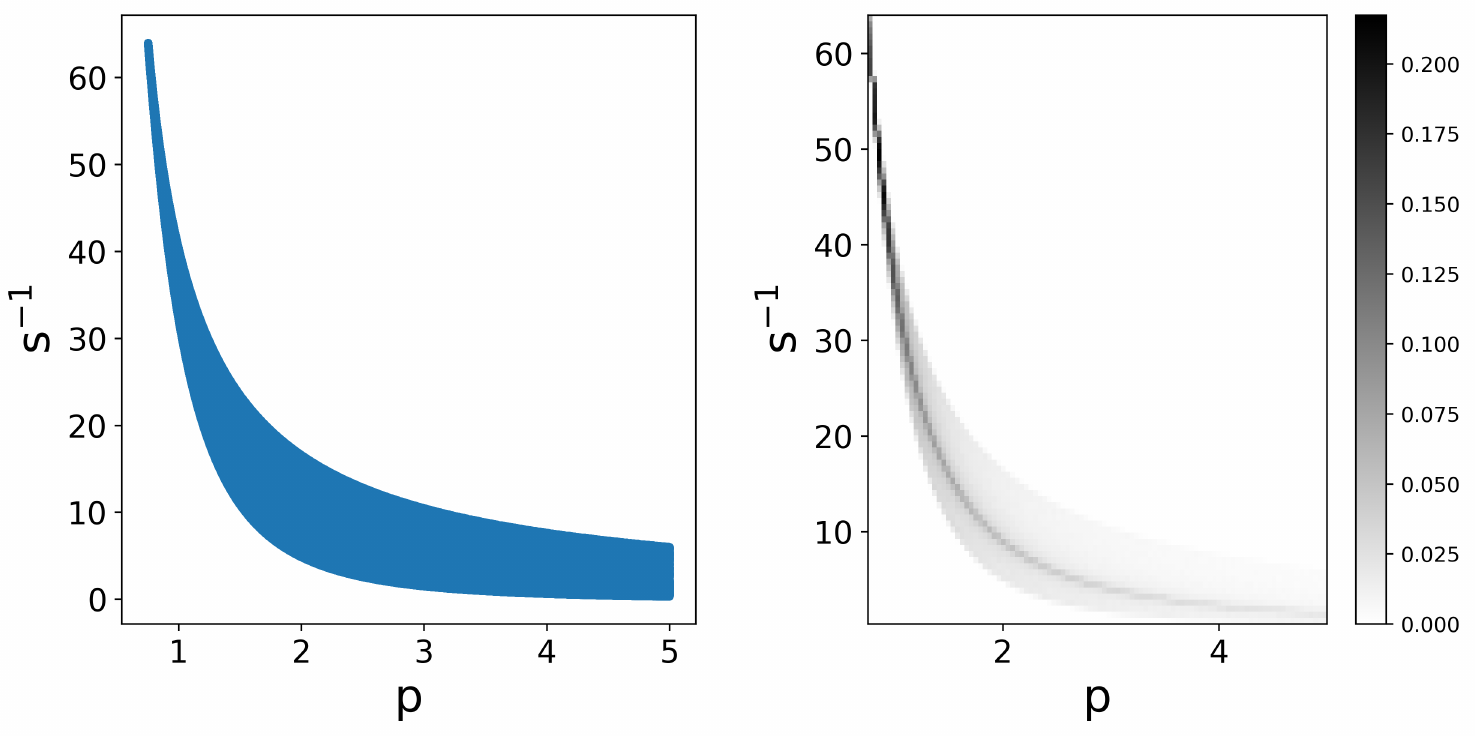}
    \caption{The trace-determinant region formed by uniformly
    sampling $2\cdot 10^6$ two-qubit states and using Eqs. (\ref{eq:TrIF}) and (\ref{eq:DetF}).
    The left panel illustrate\gff{s} the physical region of the plane, while 
    the right one shows the density of states. For 
    graphical reasons the range of $p$ has been limited to values below 5 
    (the trace of the inverse \gf{QFIM} is actually unbounded).}
	\label{fig:TrDetImage}
\end{figure}

As a first step, we investigate how the trace of the inverse and the 
determinant of the %QFI
\gf{QFIM} are related by sampling numerically
random states. The results are presented in
%Figure
\gf{Fig.} \ref{fig:TrDetImage}, where points on the determinant-trace plane
have been obtained sampling from a uniform distribution.
\par
The first observation is that the sloppiness lies in the range 
$s\in [\frac1{64}, \infty)$, while the range of precision 
is $p\in [\frac34,\infty)$. We also see that there are probe states 
leading to high values of the determinant 
(minimum sloppniness) and low values of the inverse of the trace 
(better precision), i.e.,  avoiding sloppiness is not necessarily 
decreasing precision. Indeed, the optimal point at $\left(\frac{3}{4},\frac1{64}\right)$ optimize\gff{s} both quantities at the same time, also saturating the bound in Eq. (\ref{matb}).
\par
To find the probe states that achieve that optimal condition, 
we can either minimize $p$ or $s$, since 
optimal states optimize both. To this aim, we rewrite 
Eq. (\ref{eq:DetF}) as follows
\begin{equation}
    \begin{split}
        \text{Det}[Q] = 1024 &\Big[\left[\alpha ^2+\delta ^2\right]^2-2 \alpha ^2 \delta ^2 \left[1+\cos \left(2 \phi _{\delta }\right)\right]\Big]\times\\
        &\Big[\left[\beta ^2+\gamma ^2\right]^2-2 \beta ^2 \gamma ^2 \left[1+\cos \left(2 (\phi _{\beta }- \phi _{\gamma })\right)\right]\Big].
    \end{split}
    \label{eq:det1}
\end{equation}
The expression consists of two factors, each dependent on a
different phase. In order to maximize the expression we can
maximize the two factors using suitable phases. The last term 
in both factors is always negative, thus the maximum it
can assume is $0$, when $\cos(x) = -1$. Alternatively, it may
also be that either $\alpha$ or $\delta$, and either $\beta$ or
$\gamma$ are $0$. These two options characterize two classes
of solutions. Having cancelled the negative part, the terms
$(\alpha^2+\delta^2)$ and $(\beta^2+\gamma^2)$ remains. Given this, 
a good change of variable that accounts for normalization is given by
imposing $\alpha^2+\delta^2 = \sin^2\theta$ and
$\beta^2+\gamma^2 = \cos^2\theta$. By doing so we get
\begin{equation}
    \text{Det}[Q] = 64 \sin^4 2\theta\,.
\end{equation}
In this way, we confirm the numerical evidence and show that we have optimal states for $\sin 2\theta =1$. 
The optimal states are thus given by the two sets
\begin{align}
    \ket{\psi_{\text{sep}}} & \in \Bigg\{
        \begin{pmatrix}\frac{1}{\sqrt{2}}\\\frac{1}{\sqrt{2}}e^{i\phi}\\0\\0\end{pmatrix},
        \begin{pmatrix}\frac{1}{\sqrt{2}}\\0\\\frac{1}{\sqrt{2}}e^{i\phi}\\0\end{pmatrix},
        \begin{pmatrix}0\\\frac{1}{\sqrt{2}}e^{i\phi}\\0\\\frac{1}{\sqrt{2}}\end{pmatrix},
        \begin{pmatrix}0\\0\\\frac{1}{\sqrt{2}}e^{i\phi}\\\frac{1}{\sqrt{2}}\end{pmatrix}
        \Bigg\}
        \label{eq:opt_factorizable}   
\\
    \ket{\psi_{\text{ent}}} & =    \left(
        \alpha, \beta e^{i\phi}, \pm i\sqrt{\frac{1}{2}-\beta^2}e^{i\phi}, \pm i\sqrt{\frac{1}{2}-\alpha^2}
    \right)^T
    \label{eq:opt_entangled}
\end{align}
with $\alpha,\beta \in [0,\frac{1}{\sqrt{2}}]$.
The first class of solutions is made of factorized states, which are %\gff{basically}
reducible to opportune combinations of the first %qbit
\gf{qubit} being in
a superposition of 0 and 1 with equal probability, and the second one being
either $\ket{1}$ or $\ket{0}$.
The second class of solutions spans all possible concurrences
from $0$ to $1$. We can see that the first class is actually 
contained in the second. Nevertheless, it is useful to keep 
them separate to study the effect of entanglement.
Notice that the trace of the inverse %QFI
\gf{QFIM} may be written as
$p = 3/(4 \sin^2 2\theta)$,
confirming that the optimal states in Eqs. (\ref{eq:opt_factorizable}) and (\ref{eq:opt_entangled}), also minimize $p$.
%%%
\subsection{Characterization of optimal states}
\label{ch:characterization}
We now proceed to ask if it is possible to more easily 
characterize optimal probe states. To this aim, it is useful to 
look again at $U$ 
in Eq. (\ref{eq:U}). %noticing that it is a block matrix, i.e., is mixing the
%upper and lower components, and the same may be seen from Eq. (\ref{eq:finState}).
%Moreover, the two blocks in %Equation \ref{eq:Ublock}
%\gf{Eq. \ref{eq:Ublock}} share the similar form
\gf{We can see that if we block diagonalize $U$, we get 2 blocks of the form}
\begin{equation}
    M = e^{i\phi} \begin{pmatrix}
        \cos \theta & -i\sin\theta\\
        -i\sin \theta & \cos\theta
    \end{pmatrix}
    \,,
    \label{eq:M}
\end{equation}
i.e. represents a rotation $M = e^{i\phi}e^{-i\theta\sigma_x}$ 
along the $x$ axis on the Bloch sphere of an abstract \gf{qubit}. %qbit.
Therefore our $U$ acts on those
subspaces as a rotation, and the most sensible
states are living on the $y-z$ plane of the corresponding Bloch sphere.
With this observation in mind, we can now easily generate all optimal
states, taking for example the state $\ket{0}$ and rotating it along
the $x$ axis with $e^{i\theta\sigma_x}$
\begin{equation}
    R_x(\theta)\ket{0}=
    e^{i\theta\sigma_x}\begin{pmatrix}1\\0\end{pmatrix} =
    \left(\begin{array}{c}
    \cos (\theta ) \\
    \gf{-} i \sin (\theta ) \\
    \end{array}
    \right) = 
    \left(\begin{array}{c}
        x \\
        \pm i \sqrt{1-x^2} \\
        \end{array}
        \right)
\end{equation}
Taking this for the pairs $\alpha\sim\delta$ and $\beta\sim\gamma$,
by combining them we get
\begin{equation}
    \ket{\psi_{\text{ent}}} =     \left(
        \alpha, \beta e^{i\phi}, \pm i\sqrt{\frac{1}{2}-\beta^2}e^{i\phi}, \pm i\sqrt{\frac{1}{2}-\alpha^2}
    \right)^T
\end{equation}
which we recognize to describe exactly (all) the states in Eq. (\ref{eq:opt_entangled}).
%%%
\subsection{Entanglement is not a resource}
Having found optimal solutions, we can now better 
assess whether entanglement play\gff{s} any role in determining optimality
of the probe states. To this aim, let us reconsider the data shown in Fig. \ref{fig:TrDetImage}, adding an equal number of factorizable states (orange). 
\begin{figure}[h!]
	\centering\includegraphics[width=0.75\textwidth]{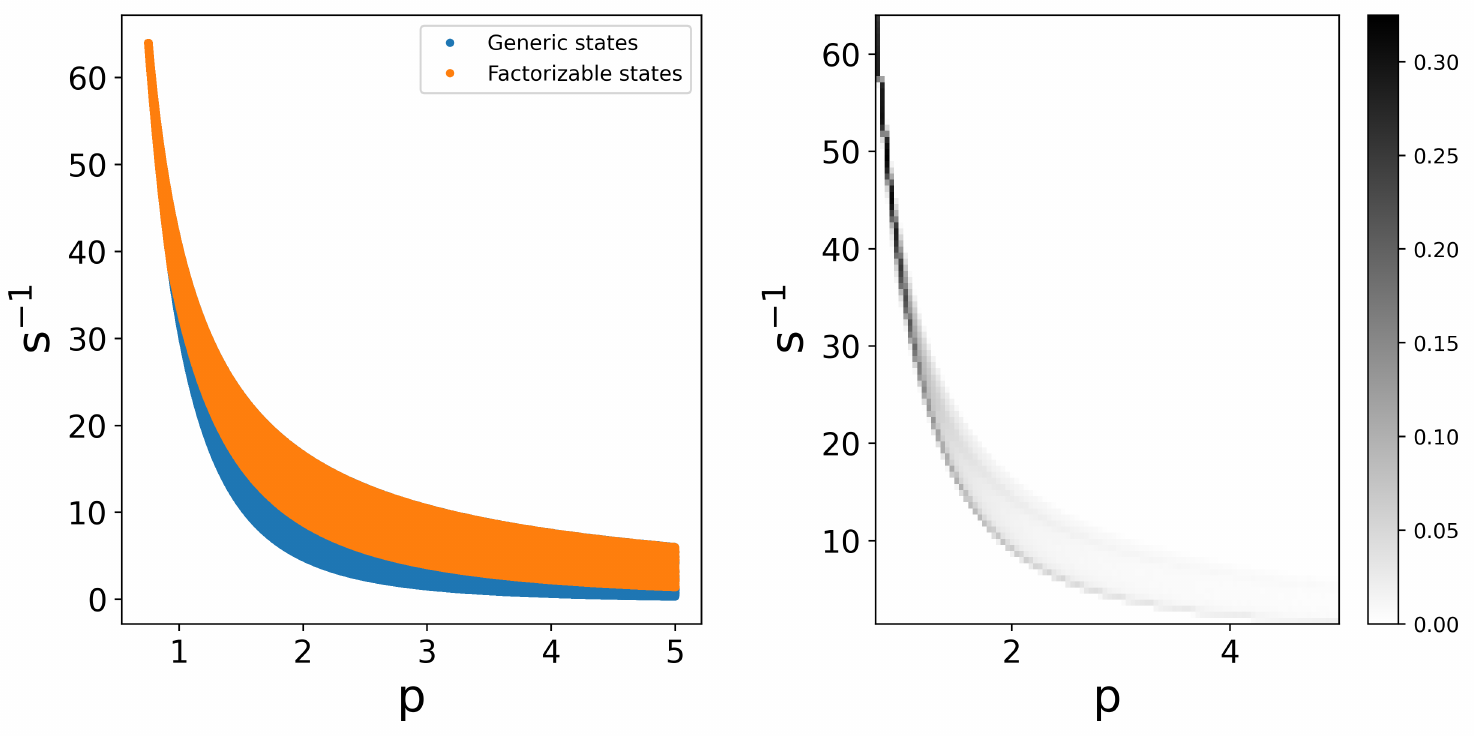}
	\caption{The left panel shows the distribution of states in the 
	trace-determinant plane formed by uniformly sampling $2\cdot 10^6$ 
	states (blue, no control on whether they are factorized or not), 
	and $2\cdot 10^6$  factorizable states (orange). The right panel 
	shows the density of factorizable states on the same plane.
    For graphical reasons the range of $p$ has been limited to values
    below 5 (actually the trace of the inverse QFIM is unbounded). }
	\label{fig:FactorizableTrDetImage}
\end{figure}

\begin{figure}[h!]
	\centering\includegraphics[width=0.75\textwidth]{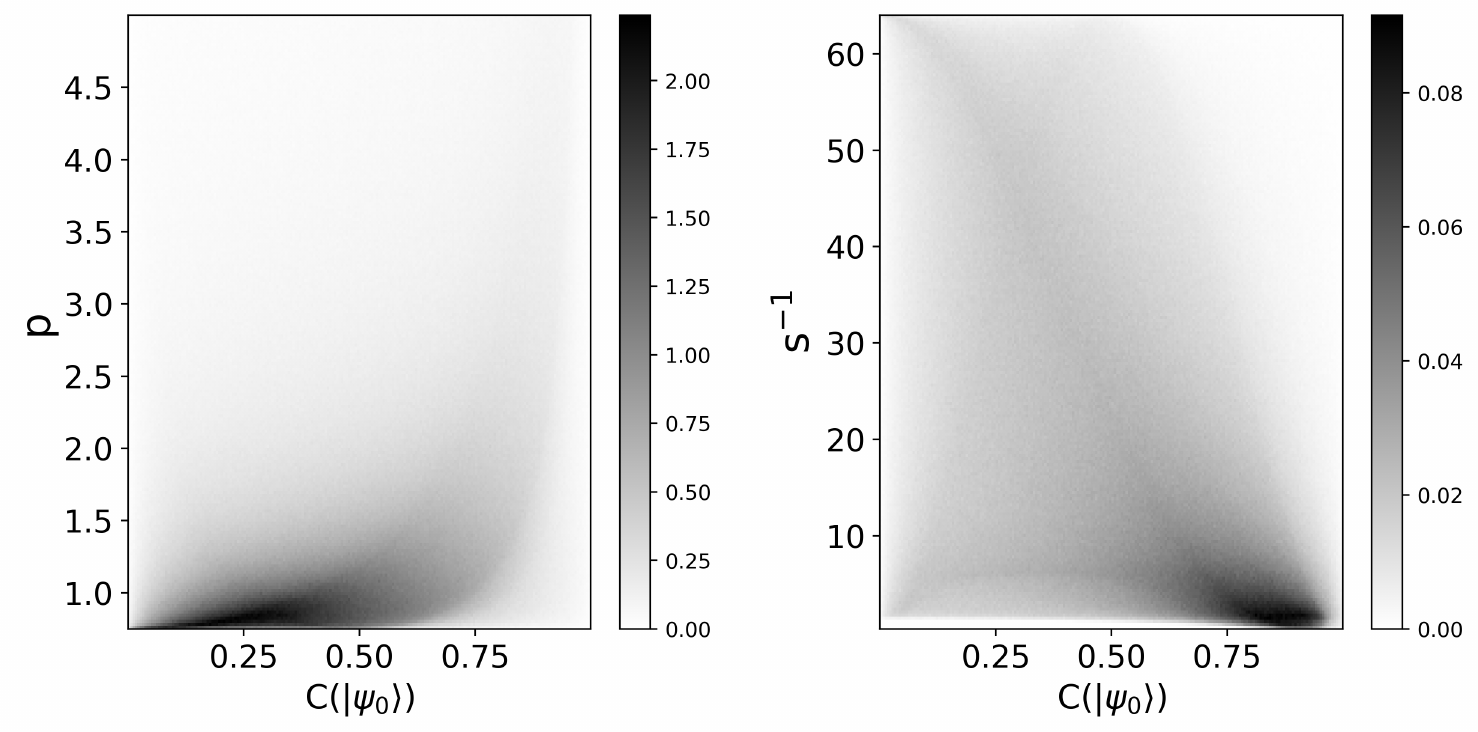}
	\caption{The left and right panels show the precision $p$ and 
	the inverse of sloppiness 
	$s^{-1}$ as a function of 
	the concurrence of the probe state, respectively.
    The graphs are density maps built by sampling %$2 \cdot 10^6$ 
    \gf{$2 \cdot 10^7$} randomly
    generated states. This shows that precision and sloppiness 
    are not related to the degree of entanglement of the probe state.}
	\label{fig:trdetconc}
\end{figure}

Results are shown in Fig. \ref{fig:FactorizableTrDetImage}. 
What we see is that, despite not having anymore the lower part, we still get
pretty much the same image and, once again, we have a higher density
of states along the same curve. To complete the analysis we show in Fig. \ref{fig:trdetconc} how the trace of the inverse and the determinant of 
the %QFI
\gf{QFIM} are related to the concurrence of the probe. 
The plot shows that maximum precision is mostly 
achieved by probe states with small concurrence, while maximum sloppiness 
is mostly achieved by probes with 
large concurrence. Since every point in the 
image is reached by at least a state, it is clear
that there is no relationship between optimality and 
entanglement.

%%%

% \subsection{Suboptimal solutions at fixed trace}
% \subsection{Optimization at fixed trace}

\subsection{Probes that minimize sloppiness 
at fixed precision}
\label{sec:belloptimization}
\begin{figure}[h!]
	\centering\includegraphics[width=1\textwidth]{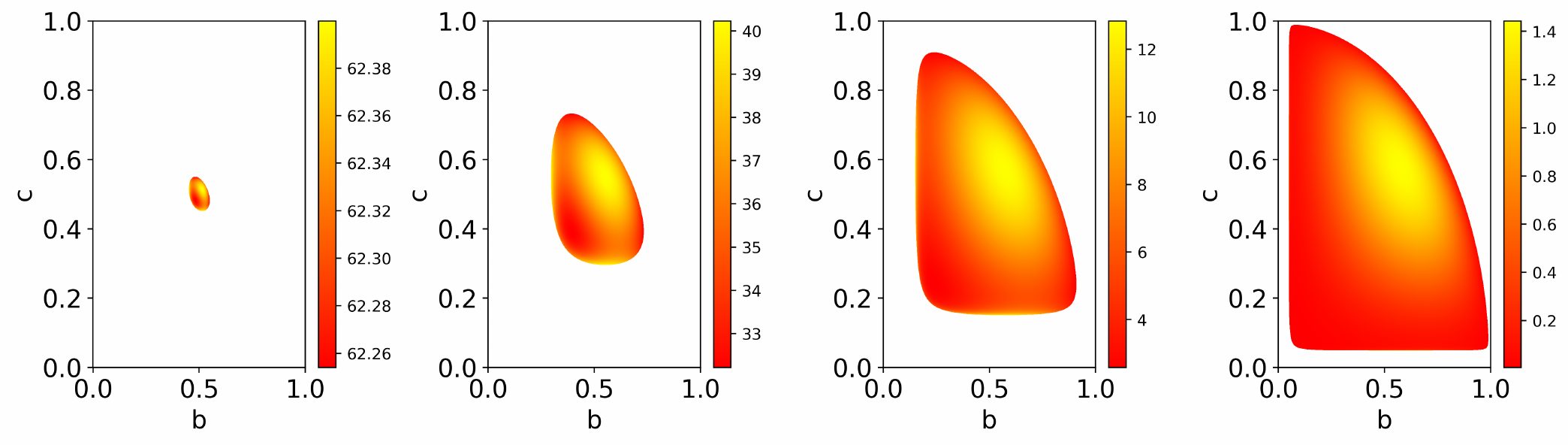}
    \caption{The four panels show the determinant of the %QFI
    \gf{QFIM} as a function of $b$ and $c$ %. The value of the precision parameter $p$ is set to 1. Right panel: 
    at different values of $p$. They are at respectively $p=0.76, 1, 2.5, 20$.}
    %twodimensional heatmap of the same quantity.}
	\label{fig:Detbct}
\end{figure}

\gf{Looking back at Fig. \ref{fig:TrDetImage}, we now want to characterize
the states lying on the top curve, i.e., states minimizing sloppiness 
for a given level of precision (i.e., fixed $p$).
We do this \gff{by} rewriting the determinant as a function of $p$ and maximizing that function over the other parameters. The main problem in doing this in the standard basis is that we get complex equations. The derivation can be greatly simplified by switching to the Bell basis. We start by expressing
the QFIM and the optimal states in the Bell basis.}
For convenience let's rewrite $U$ in the Bell basis
\begin{equation}
%	U = \hbox{Diag} 	
%	\left\{
%		 e^{-i (\lambda_1-\lambda_2+\lambda_3)} , e^{-i (\lambda_1+\lambda_2-\lambda_3)}, e^{i (\lambda_1-\lambda_2-\lambda_3)} , e^{i (\lambda_1+\lambda_2+\lambda_3)}		\right\}.
    \gf{U = \hbox{Diag} 	
	\left\{
		 e^{-i (\lambda_1-\lambda_2+\lambda_3)} ,  e^{i (\lambda_1-\lambda_2-\lambda_3)} , e^{-i (\lambda_1+\lambda_2-\lambda_3)}, e^{i (\lambda_1+\lambda_2+\lambda_3)}		\right\}.}
\end{equation}
%and use
\gf{We use} latin letters to denote the amplitudes of %the probe state 
\gf{a generic probe state} in the %same
\gf{Bell} basis $\ket{\psi_0} = (a, b  e^{i \phi _{b }}, c  e^{i \phi _{c }},d  e^{i \phi _{d }}  )^T$, \gf{properly normalized}. The evolved state after the application of 
$U$, is given by 
\begin{equation}
%    \ket{\psi_{\underline{\lambda}}} = \left(
%    \begin{array}{c}
%    a e^{-i (\lambda_1-\lambda_2+\lambda_3)} \\
%    b e^{i \left(\phi _b-\lambda_1-\lambda_2+\lambda_3\right)} \\
%   c e^{i \left(\phi _c+\lambda_1-\lambda_2-\lambda_3\right)} \\
%    d e^{i \left(\phi _d+\lambda_1+\lambda_2+\lambda_3\right)} \\
%    \end{array}
%    \right)\,,
    \gf{\ket{\psi_{\underline{\lambda}}} = \left(
    \begin{array}{c}
    a\, e^{-i (\lambda_1-\lambda_2+\lambda_3)} \\
    b\, e^{i \left(\phi _b+\lambda_1-\lambda_2-\lambda_3\right)} \\
    c\, e^{i \left(\phi _c-\lambda_1-\lambda_2+\lambda_3\right)} \\
    d\, e^{i \left(\phi _d+\lambda_1+\lambda_2+\lambda_3\right)} \\
    \end{array}
    \right)\,,}
\end{equation}
and the elements of the %QFI
\gf{QFIM} %\gff{matrix}
by
\gf{
\begin{equation}
    \begin{split}
        &Q_{11} = 16 \left(1-b^2-d^2\right) \left(b^2+d^2\right)\\
        &Q_{22} = 16 \left(1-b^2-c^2\right) \left(b^2+c^2\right)\\
        &Q_{33} = 16 \left(1-c^2-d^2\right) \left(c^2+d^2\right)\\
        &Q_{12} = Q_{21} = 16 \left(b^4+b^2 \left(c^2+d^2-1\right)+c^2 d^2\right)\\
        &Q_{13} = Q_{31} = 16 \left(d^2-\left(b^2+d^2\right) \left(c^2+d^2\right)\right)\\
        &Q_{23} = Q_{32} = 16 \left(c^2 \left(b^2+d^2-1\right)+b^2 d^2+c^4\right)
    \end{split}
    \label{eq:bellQFI}
\end{equation}
}
which depend only on the real parameters $b,c,d$ and not on the phases. 
Solutions will then be defined up to relative phases applicable at the end.
The precision $p$ and the determinant of the %QFI
\gf{QFIM} %\gff{matrix}
may be expressed as 
\begin{align}
    p & = \frac{3}{64} \left(\frac{1}{b^2}+\frac{1}{c^2}+\frac{1}{d^2}+\frac{1}{1-b^2-c^2-d^2}\right) \\ 
    \frac1s & = 16384 \left(1-b^2-c^2-d^2\right)^2 b^2 c^2 d^2\,,
\end{align}
from which it is straightforward to obtain the optimal states
\begin{equation}
%    \ket{\psi_{\text{opt}}} = \left(
%        \begin{array}{c}
%        \frac{1}{2}  \\
%        \frac{1}{2}  e^{i \phi _{\beta }} \\
%        \frac{1}{2}  e^{i \phi _{\gamma }} \\
%        \frac{1}{2}  e^{i \phi _{\delta }} \\
%        \end{array}
%        \right).
% should use latin letters for Bell's basis
    \gf{\ket{\psi_{\text{opt}}} =   \frac{1}{2} \left(
      1 ,
          e^{i \phi _{b }} ,
         e^{i \phi _{c }},       
         e^{i \phi _{d }}         \right)^T\,.}
\end{equation}
%We can now restart from where we stopped in Section 
%\ref{sec:suboptimization} and find the states that maximize the 
%determinant at fixed precision.
We now begin to look for optimal states at fixed precision. We express the amplitude $d\equiv d(b,c,p)$ in terms of precision arriving at
\begin{equation}
   \frac1s = \frac{49152 b^4 c^4 \left(1-b^2-c^2\right)}{b^2 \left(64\, c^2 p-3\right)-3 c^2}.
   \label{eq:det44}
\end{equation}
The situation is illustrated in Fig. \ref{fig:Detbct}, where we show
the determinant as a function of $b$ and $c$  at fixed $p$,
taking into account the constraints on the domain.

\mga{The graphs show similar features for all values of $p$, and are characterized by 3 maxima. One is on the diagonal and the other two are on the edges of the domain (barely visible in the plots), sharing symmetrically one of the two components with the diagonal one. As a matter of fact,  the function becomes more and more steep on the edges, making the yellow regions to nearly disappear, though the maxima are still there.}
%There are 3 maxima, achieving the same value of the determinant. Analogue 
%plots may be obtained for any value of $p$. 
%For the maxima on the diagonal $b=c$ we have 
\gf{Substituting a maximum in Eq. (\ref{eq:det44}) we get}
% \begin{equation}
%     \frac1s= \frac{\left(\sqrt{(p-\frac3{16})(p-\frac34)}-16 p+3\right) \left(\sqrt{(p-\frac3{16})(p-\frac34)}+8 p+3\right)^3}{3-36 p^4 \left(2 \sqrt{(p-\frac3{16})(p-\frac34)}-16 p\right)},
% \end{equation}
% wrong equation
%    \frac1s= \frac{\left(8 p+8 \sqrt{\left(p-\frac{3}{4}\right)
%   \left(p-\frac{3}{16}\right)}+3\right)^3 \left(16 p-8
%   \sqrt{\left(p-\frac{3}{4}\right)
%   \left(p-\frac{3}{16}\right)}-3\right)}{36 p^4 \left(16 p+16
%   \sqrt{\left(p-\frac{3}{4}\right)
%   \left(p-\frac{3}{16}\right)}-3\right)}
% more compact form
\gf{
\begin{equation}
\frac1s= \frac{\left(8 p-8 \sqrt{\left(p-\frac{3}{4}\right) \left(p-\frac{3}{16}\right)}-3\right) \left(8 p+8 \sqrt{\left(p-\frac{3}{4}\right)
   \left(p-\frac{3}{16}\right)}+3\right)^3}{108 p^4}
\end{equation}
}
which expresses the minimum sloppiness achievable for a given precision $p$. \mga{To the leading around the minimum value of $p$, we have $1/s = 64 - 512/3 (p - 3/4) + O(p - 3/4)^2$}.
The \gff{corresponding} (sub)optimal states, achieving those values of precision and sloppiness are given by
\begin{equation}
    \ket{\psi_{\text{sub}}} = \left(        \kappa_2,
        \kappa_1 e^{i \phi_b},
        \kappa_1 e^{i \phi_c},
        \kappa_1 e^{i \phi_d}\right)^T
        \end{equation}
\gf{with
\begin{align}
    \kappa_1 &= \frac{1}{4}\sqrt{\frac{8 \sqrt{\left(p-\frac{3}{4}\right)
   \left(p-\frac{3}{16}\right)}+8p+3}{3 p}} \\ \kappa_2 & = \frac{1}{4}\sqrt{\frac{-8 \sqrt{\left(p-\frac{3}{4}\right)
   \left(p-\frac{3}{16}\right)}+8p-3}{p}}
\end{align}
}
where $\kappa_2$ can be switched in either of the four components.
\section{Robustness against noise}
\label{s:noise}
\begin{figure}[h!]
    \centering\begin{minipage}{0.47\textwidth}\includegraphics[width=1\textwidth]{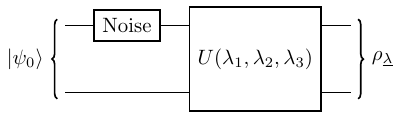}
	%\caption{Quantum circuit representing how noisy channel %alter
    %the prepared probe states before performing the $U$ %evolution.}
    \end{minipage}
    \hfill
	\centering\begin{minipage}{0.47\textwidth}\includegraphics[width=1\textwidth]{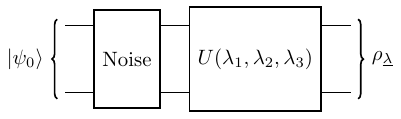}
	%\caption{Quantum circuit representing how noisy channel alter
    %the prepared probe states before performing the $U$ evolution.}
    \end{minipage}
    \caption{Quantum circuits representing how noisy channels alter
    the probe states before performing the $U$ evolution. \gf{We consider the action of the noisy channel on just one channel(left panel), and on both(right panel).}}
	\label{fig:NoisyChannel}
\end{figure}

Having found an entire class of optimal states, we now want to characterize
their robustness against noise in the preparation stage.
The %circuit
\gf{circuits} under consideration %is shown
are shown in Fig. \ref{fig:NoisyChannel}. We assume 
that after the preparation of the probe state %the system
\gf{one, or both channels} may be %subjected
\gf{subject} to some form of noise before entering the gate.
At this stage, we avoid classifying the states based on specific 
noise models, as these could be tied to particular implementations 
of the system \cite{rossi2014engineering}. 
Instead, our focus is on assessing the robustness 
and performance of the solutions obtained. To this aim, we apply 
some simple noise models and examine how optimal precision changes 
under these conditions. For simplicity, we have selected three 
distinct classes of probe states to illustrate how different 
\gf{optimal} states respond 
to noise.  

\begin{figure}[h!]
\centering\includegraphics[width=0.75\textwidth]{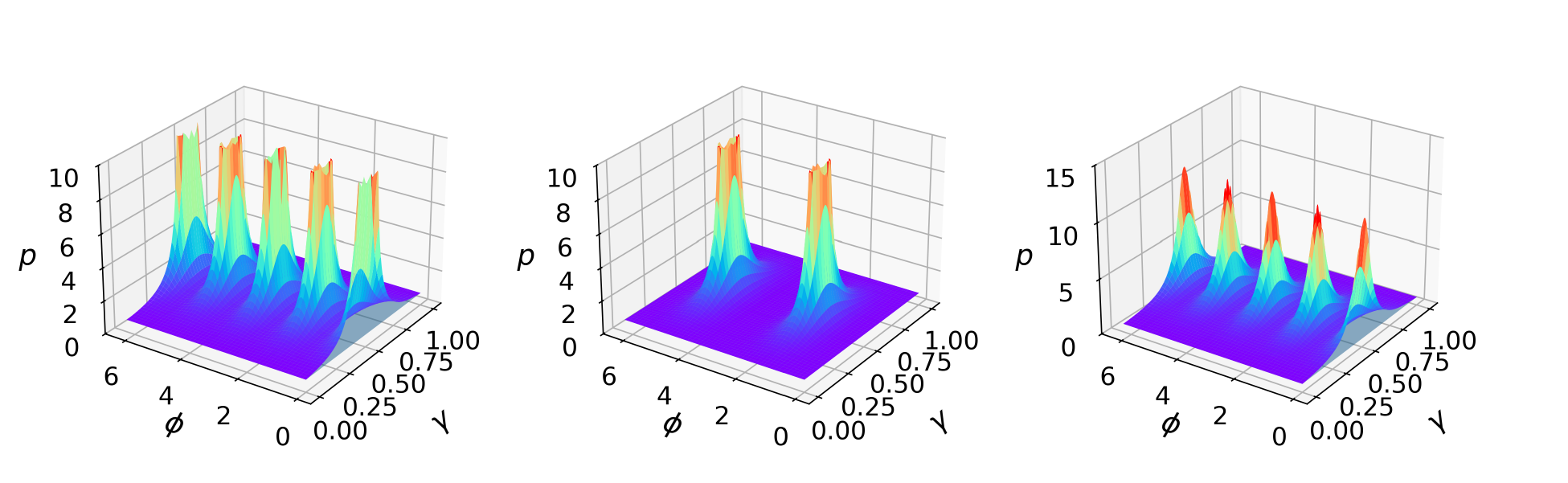}
\centering\includegraphics[width=0.75\textwidth]{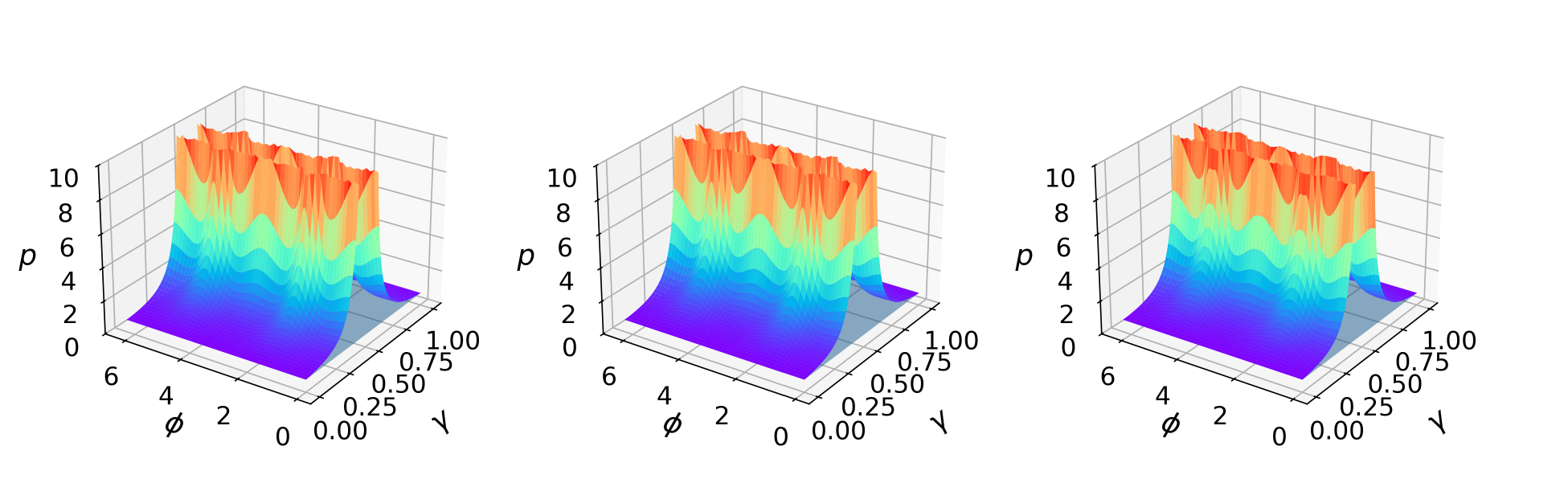}	
	\caption{The precision $p$ as a function of the bit-flip noise parameter $\gamma$ and state parameter $\phi$. The upper panels illustrate results for the noise channel applied only on the second qubit, and the 
	lower ones for the noise on both channels. The three columns 
	are for the three classes of states in Eq. (\ref{eq:classes}).}
	\label{fig:bitFlip}
\end{figure}

\begin{figure}[h!]
\centering\includegraphics[width=0.75\textwidth]{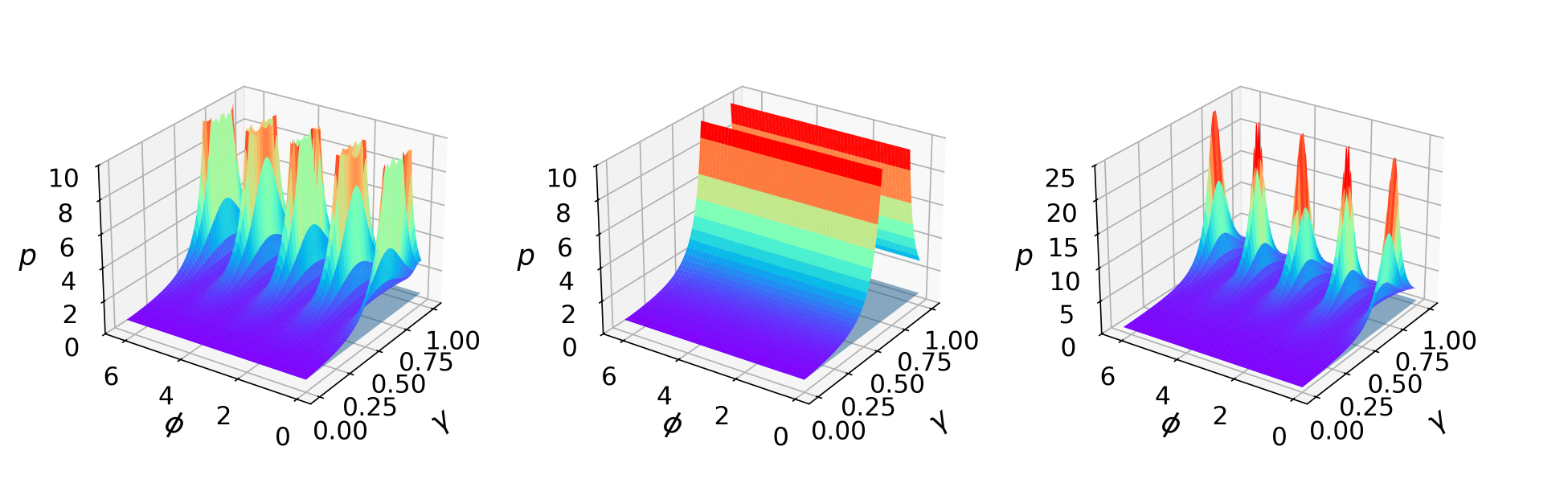}
\centering\includegraphics[width=0.75\textwidth]{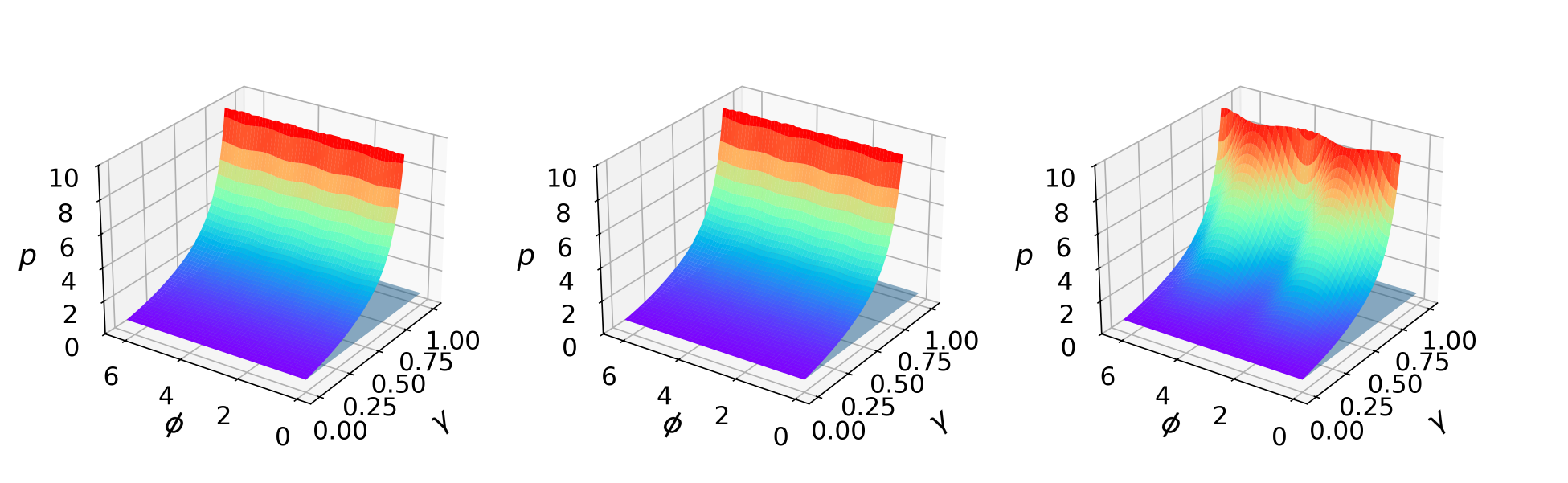}	
	\caption{The precision $p$ as a function of the depolarizing noise parameter $\gamma$ and state parameter $\phi$. The upper panels illustrate results for the noise channel applied only on the second qubit, and the 
	lower ones for the noise on both channels. The three columns 
	are for the three classes of states in Eq. (\ref{eq:classes}).}
	\label{fig:depolarizing}
\end{figure}

These classes are the following
%\begin{align}
 %       \ket{\psi_1} & \equiv \begin{pmatrix}\frac{1}{\sqrt{2}}\\\frac{1}{\sqrt{2}}e^{i \phi}\\0\\0\end{pmatrix} = \ket{0}\otimes\begin{pmatrix}\frac{1}{\sqrt{2}}\\\frac{1}{\sqrt{2}}e^{i \phi}\end{pmatrix}\qquad
%        \ket{\psi_2} \equiv \begin{pmatrix}\frac{1}{\sqrt{2}}\\0\\\frac{1}{\sqrt{2}}e^{i \phi}\\0\end{pmatrix} = \begin{pmatrix}\frac{1}{\sqrt{2}}\\\frac{1}{\sqrt{2}}e^{i\phi}\end{pmatrix}\otimes \ket{0}\\
%        \ket{\psi_3} & \equiv \begin{pmatrix}0.3\\0.2\ e^{i\phi}\\i\sqrt{\frac{1}{2}-0.2^2}e^{i\phi}\\i\sqrt{\frac{1}{2}-0.3^2}\end{pmatrix}.
%    \label{eq:classes}
%\end{align}
\begin{equation}
    \begin{split}
        \ket{\psi_1} & \equiv \begin{pmatrix}\frac{1}{\sqrt{2}}\\\frac{1}{\sqrt{2}}e^{i \phi}\\0\\0\end{pmatrix} = \ket{0}\otimes\begin{pmatrix}\frac{1}{\sqrt{2}}\\\frac{1}{\sqrt{2}}e^{i \phi}\end{pmatrix}\qquad
        \ket{\psi_2} \equiv \begin{pmatrix}\frac{1}{\sqrt{2}}\\0\\\frac{1}{\sqrt{2}}e^{i \phi}\\0\end{pmatrix} = \begin{pmatrix}\frac{1}{\sqrt{2}}\\\frac{1}{\sqrt{2}}e^{i\phi}\end{pmatrix}\otimes \ket{0}\\
        \ket{\psi_3} & \equiv \begin{pmatrix}0.3\\0.2\ e^{i\phi}\\i\sqrt{\frac{1}{2}-0.2^2}e^{i\phi}\\i\sqrt{\frac{1}{2}-0.3^2}\end{pmatrix}.
    \label{eq:classes}
    \end{split}
\end{equation}
In particular, %we investigate the effect of noise acting on one of the two %qbits
%\gf{qubits}, and
\gf{we} consider the action of 1) bit flip noise, where the %state entering the gate
\gf{final state} is given by
\begin{equation}
    \rho_{\underline{\lambda}} = (1-\gamma) U\rho_0 U^\dagger + \gamma U (\sigma_x\otimes\mathbb{I}) \rho_0 (\sigma_x\otimes\mathbb{I})^\dagger U^\dagger  \qquad 0\leq\gamma\leq 1
    \label{eq:noiseBF1}
\end{equation}
\gf{if it acts only on one channel, and}
\begin{equation}
\gf{
\begin{split}
    \rho_{\underline{\lambda}} &= (1-2\gamma(1-\gamma)-\gamma^2) U\rho_0 U^\dagger + + \gamma(1-\gamma) U (\sigma_x\otimes\mathbb{I}) \rho_0 (\sigma_x\otimes\mathbb{I})^\dagger U^\dagger\\
    &+ \gamma(1-\gamma) U (\mathbb{I}\otimes\sigma_x) \rho_0 (\mathbb{I}\otimes\sigma_x)^\dagger U^\dagger
    + \gamma^2 U (\sigma_x\otimes\sigma_x) \rho_0 (\sigma_x\otimes\sigma_x)^\dagger U^\dagger 
\end{split}
    }
    \label{eq:noiseBF2}
\end{equation}
%with $E_k=\sigma_k \otimes \mathbb{I} $, 
\gf{if it acts on both}, and 2) the depolarizing noise, 
%where the state entering the gate is given by  
\gf{with the final state given by}
\begin{equation}
%    \rho_{\underline{\lambda}} = (1-\gamma)U \rho_0 U^\dagger + \frac{\gamma}{3}\sum\limits_{k \in \{x,y,z\}} U E_k^\dagger \rho_0 E_k U^\dagger 
\gf{    \rho_{\underline{\lambda}} = (1-\gamma)U \rho_0 U^\dagger + \frac{\gamma}{3}\sum\limits_{k \in \{x,y,z\}} U (\sigma_k\otimes\mathbb{I}) \rho_0 (\sigma_k\otimes\mathbb{I})^\dagger U^\dagger
}
\end{equation}
\gf{if it acts only on one channel, and}
\begin{align}
\rho_{\underline{\lambda}} &= (1-\gamma)^2U \rho_0 U^\dagger + (1-\gamma)\frac{\gamma}{3}\sum\limits_{k} U (\sigma_k\otimes\mathbb{I}) \rho_0 (\sigma_k\otimes\mathbb{I})^\dagger U^\dagger\notag \\
& + (1-\gamma)\frac{\gamma}{3}\sum\limits_{l} U (\mathbb{I}\otimes\sigma_l) \rho_0 (\mathbb{I}\otimes\sigma_l)^\dagger U^\dagger + \frac{\gamma^2}{9}\sum\limits_{k,l} U (\sigma_k\otimes\sigma_l) \rho_0 (\sigma_k\otimes\sigma_l)^\dagger U^\dagger
\end{align}
\gf{if it acts on both ($k,l \in \{x,y,z\}$).}
The precision %may be
\gf{is} calculated using Eq. (\ref{eq:generalQFI}), and 
the overall goal is to see whether there exist probe states that maintain their sensitivity in the presence of noise. 
Results are summarized in Figs. \ref{fig:bitFlip} \gf{and \ref{fig:depolarizing}}, where we show precision as a function of the noise parameter 
$\gamma$ and the state parameter $\phi$ for the three classes of states 
defined in Eq. (\ref{eq:classes}).

%%%
What we can see from Figs. \ref{fig:bitFlip} and \ref{fig:depolarizing} is that 
\gf{in all cases the function 
is convex for small $\gamma$ and thus robustness is ensured}. \gf{Few additional remarks are in order: The first is that in the case of \gff{a} single channel, for both errors there are phases for which the precision is comparable to the optimal one for every $\gamma$. \mga{
This phases happens to be the same for $ |\psi_1\rangle$ and $ |\psi_3\rangle$ in the two cases shown, but in general they are error dependent. In fact,  $ |\psi_2\rangle$ shows some robust phases for the bit-flip noise, but none in the depolarizing case.} Looking again at the single channel case, we can also see that the third state has bounded precision for any choice of $\phi$ and $\gamma$ for both kind\gff{s} of errors. Lastly, we see how in the two channel case there are values of $\gamma$ for which the \mga{bound to} precision diverges for all phases, for both error models.} 
\mga{Notice also that in the limit $\gamma=1$ the noise channel corresponds to applying a unitary bit flip with unit probability, such that the pure input state gets mapped onto another optimal pure state. In this case, the bound to precision $p$ returns back to the minimum value $p=3/4$. This observation also explains the symmetry of the plot around the value $\gamma=1/2$.}
%%%%
%%%%%%%%%%%%%%%%%%%%%%%%%%%%%%%%%%%%%%%%%%
\section{Discussion and conclusions}
\label{s:outro}
We have addressed the characterization of two-qubit gates, focusing on precision bounds in the joint estimation of the three parameters defining their Cartan decomposition. By considering a generic probe state at the input of the Cartan kernel, we identified classes of optimal probe states that maximize precision while minimizing sloppiness. For these probes, quantum incompatibility vanishes, allowing the three parameters to be estimated without any additional quantum noise. In addition, we have thoroughly analyzed the properties of these optimal probes and assessed their robustness against noise, revealing a subset of states that are more resilient to noise than others. 

The optimal probes allow one to achieve a precision bound of $p=3/4$, 
meaning that the precision of any multidimensional set of estimators 
is constrained by $\sum_{j=1}^3 \hbox{Var}\lambda_j \geq 3/4$. Furthermore, the same set of probes ensures the minimum sloppiness, $s=1/\text{Det}[Q] 
= 1/64$.  We have also determined the optimal precision-sloppiness tradeoff, providing the minimum value of $s$ for a fixed level of precision $p$. 
This tradeoff is given by:
\gf{
\begin{equation}
s = \frac{108 p^4}{\left(8 p-8 \sqrt{\left(p-\frac{3}{4}\right) \left(p-\frac{3}{16}\right)}-3\right) \left(8 p+8 \sqrt{\left(p-\frac{3}{4}\right)
   \left(p-\frac{3}{16}\right)}+3\right)^3}
\end{equation}
}
with $p \in [3/4,+\infty)$

In conclusion, we have derived the optimal probe states to characterize 
the Cartan kernel of a generic two-qubit \gf{gate}. These 
probes %allows
\gff{allow} one to achieve maximum precision while eliminating quantum incompatibility. This enables noiseless, multidimensional, parameter 
estimation. \gf{These findings set a new benchmark for the precise characterization of two-qubit quantum gates, and shed light on the precision-sloppiness tradeoff in qubit systems.}
 %%%%%%%%%%%%%%%%%%%%%%%%%%%%%%%%%%%%%%%%%%
\section*{Acknowledgments}
This work has been partially supported by MUR via PRIN-2022 project RISQUE (contract n. 2022T25TR3).

\bibliographystyle{elsarticle-num}
\bibliography{cartanbib}

\begin{thebibliography}{10}
\expandafter\ifx\csname url\endcsname\relax
  \def\url#1{\texttt{#1}}\fi
\expandafter\ifx\csname urlprefix\endcsname\relax\def\urlprefix{URL }\fi
\expandafter\ifx\csname href\endcsname\relax
  \def\href#1#2{#2} \def\path#1{#1}\fi

\bibitem{Kraus2001}
B.~Kraus, J.~Cirac, Optimal quantum circuits for two-qubit gates, Physical
  Review A 63~(6) (2001) 062309.
\newblock \href {https://doi.org/10.1103/PhysRevA.63.062309}
  {\path{doi:10.1103/PhysRevA.63.062309}}.

\bibitem{Vidal2004}
G.~Vidal, C.~M. Dawson, Optimal quantum circuits for general two-qubit gates,
  Physical Review A 69~(1) (2004) 010301.
\newblock \href {https://doi.org/10.1103/PhysRevA.69.010301}
  {\path{doi:10.1103/PhysRevA.69.010301}}.

\bibitem{Vatan2004}
F.~Vatan, C.~Williams, Optimal realization of a generic two-qubit quantum gate,
  Physical Review A 69~(3) (2004) 032315.
\newblock \href {https://doi.org/10.1103/PhysRevA.69.032315}
  {\path{doi:10.1103/PhysRevA.69.032315}}.

\bibitem{Barenco1995}
A.~Barenco, C.~H. Bennett, R.~Cleve, D.~P. DiVincenzo, N.~Margolus, P.~Shor,
  T.~Sleator, J.~Smolin, H.~Weinfurter, Elementary gates for quantum
  computation, Physical Review A 52~(5) (1995) 3457--3467.
\newblock \href {https://doi.org/10.1103/PhysRevA.52.3457}
  {\path{doi:10.1103/PhysRevA.52.3457}}.

\bibitem{Makhlin2002}
Y.~Makhlin, Nonlocal properties of two-qubit gates and mixed states, and the
  optimization of quantum computations, Quantum Information Processing 1 (2002)
  243--252.
\newblock \href {https://doi.org/10.1023/A:1022144002391}
  {\path{doi:10.1023/A:1022144002391}}.

\bibitem{Zhang2003}
J.~Zhang, J.~Vala, S.~Sastry, K.~B. Whaley, Geometric theory of nonlocal
  two-qubit operations, Physical Review A 67~(4) (2003) 042313.
\newblock \href {https://doi.org/10.1103/PhysRevA.67.042313}
  {\path{doi:10.1103/PhysRevA.67.042313}}.

\bibitem{rezakhani2004characterization}
A.~Rezakhani, Characterization of two-qubit perfect entanglers, Physical Review
  A—Atomic, Molecular, and Optical Physics 70~(5) (2004) 052313.

\bibitem{Gilchrist2005}
A.~Gilchrist, N.~K. Langford, M.~A. Nielsen, Distance measures to compare real
  and ideal quantum processes, Physical Review A 71~(6) (2005) 062310.
\newblock \href {https://doi.org/10.1103/PhysRevA.71.062310}
  {\path{doi:10.1103/PhysRevA.71.062310}}.

\bibitem{Bennett1996}
C.~H. Bennett, D.~P. DiVincenzo, J.~A. Smolin, W.~K. Wootters, Mixed-state
  entanglement and quantum error correction, Physical Review A 54~(5) (1996)
  3824--3851.
\newblock \href {https://doi.org/10.1103/PhysRevA.54.3824}
  {\path{doi:10.1103/PhysRevA.54.3824}}.

\bibitem{Chow2012}
J.~M. Chow, J.~M. Gambetta, A.~D. Córcoles, S.~T. Merkel, J.~A. Smolin,
  C.~Rigetti, S.~Poletto, G.~A. Keefe, M.~B. Rothwell, J.~Rozen, M.~B. Ketchen,
  M.~Steffen, Universal quantum gate set approaching fault-tolerant thresholds
  with superconducting qubits, Physical Review Letters 109~(6) (2012) 060501.
\newblock \href {https://doi.org/10.1103/PhysRevLett.109.060501}
  {\path{doi:10.1103/PhysRevLett.109.060501}}.

\bibitem{Khaneja2001}
N.~Khaneja, S.~J. Glaser, Cartan decomposition of su(2) and control of spin
  systems, Chemical Physics 267~(1-3) (2001) 11--23.
\newblock \href {https://doi.org/10.1016/S0301-0104(01)00333-6}
  {\path{doi:10.1016/S0301-0104(01)00333-6}}.

\bibitem{shende2004minimal}
V.~V. Shende, I.~L. Markov, S.~S. Bullock, Minimal universal two-qubit
  controlled-not-based circuits, Physical Review A—Atomic, Molecular, and
  Optical Physics 69~(6) (2004) 062321.

\bibitem{Tucci05}
R.~R. Tucci, An introduction to cartan's kak decomposition for qc programmers,
  arXiv preprint quant-ph/0507171 (2005).

\bibitem{PhysRevA.87.012106}
C.~Sparaciari, M.~G.~A. Paris, Canonical naimark extension for generalized
  measurements involving sets of pauli quantum observables chosen at random,
  Phys. Rev. A 87 (2013) 012106.

\bibitem{Liu_2019}
J.~Liu, H.~Yuan, X.-M. Lu, X.~Wang, Quantum fisher information matrix and
  multiparameter estimation, Journal of Physics A: Mathematical and Theoretical
  53~(2) (2019) 023001.

\bibitem{Albarelli_2020}
F.~Albarelli, M.~Barbieri, M.~Genoni, I.~Gianani, A perspective on
  multiparameter quantum metrology: From theoretical tools to applications in
  quantum imaging, Physics Letters A 384~(12) (2020) 126311.

\bibitem{Razavian_2020}
S.~Razavian, M.~G.~A. Paris, M.~G. Genoni, On the quantumness of multiparameter
  estimation problems for qubit systems, Entropy 22~(11) (2020) 1197.

\bibitem{Wootters2001}
W.~K. Wootters, Entanglement of formation and concurrence., Quantum Information
  and Computation 1~(1) (2001) 27--44.

\bibitem{brida2010experimental}
G.~Brida, I.~P. Degiovanni, A.~Florio, M.~Genovese, P.~Giorda, A.~Meda, M.~G.
  Paris, A.~Shurupov, Experimental estimation of entanglement at the quantum
  limit, Physical review letters 104~(10) (2010) 100501.

\bibitem{paris2009quantum}
M.~G.~A. Paris, Quantum estimation for quantum technology, International
  Journal of Quantum Information 7~(supp01) (2009) 125--137.

\bibitem{Cramer1999}
H.~Cram{\'e}r, Mathematical methods of statistics, Vol.~26, Princeton
  university press, 1999.

\bibitem{Amari2000}
S.-i. Amari, H.~Nagaoka, Methods of information geometry, Vol. 191, American
  Mathematical Soc., 2000.

\bibitem{HOLEVO1977251}
A.~Holevo, Commutation superoperator of a state and its applications to the
  noncommutative statistics, Reports on Mathematical Physics 12~(2) (1977)
  251--271.

\bibitem{brown2003statistical}
K.~S. Brown, J.~P. Sethna, Statistical mechanical approaches to models with
  many poorly known parameters, Physical review E 68~(2) (2003) 021904.

\bibitem{brown2004statistical}
K.~S. Brown, C.~C. Hill, G.~A. Calero, C.~R. Myers, K.~H. Lee, J.~P. Sethna,
  R.~A. Cerione, The statistical mechanics of complex signaling networks: nerve
  growth factor signaling, Physical biology 1~(3) (2004) 184.

\bibitem{PhysRevLett.97.150601}
J.~J. Waterfall, F.~P. Casey, R.~N. Gutenkunst, K.~S. Brown, C.~R. Myers, P.~W.
  Brouwer, V.~Elser, J.~P. Sethna,
  \href{https://link.aps.org/doi/10.1103/PhysRevLett.97.150601}{Sloppy-model
  universality class and the vandermonde matrix}, Phys. Rev. Lett. 97 (2006)
  150601.
\newblock \href {https://doi.org/10.1103/PhysRevLett.97.150601}
  {\path{doi:10.1103/PhysRevLett.97.150601}}.
\newline\urlprefix\url{https://link.aps.org/doi/10.1103/PhysRevLett.97.150601}

\bibitem{machta2013parameter}
B.~B. Machta, R.~Chachra, M.~K. Transtrum, J.~P. Sethna, Parameter space
  compression underlies emergent theories and predictive models, Science
  342~(6158) (2013) 604--607.

\bibitem{yang2023untwining}
Y.~Yang, F.~Belliardo, V.~Giovannetti, F.~Li, Untwining multiple parameters at
  the exclusive zero-coincidence points with quantum control, New Journal of
  Physics 24~(12) (2023) 123041.

\bibitem{frigerio2024overcoming}
M.~Frigerio, M.~G. Paris, Overcoming sloppiness for enhanced metrology in
  continuous-variable quantum statistical models, arXiv preprint
  arXiv:2410.02989 (2024).

\bibitem{ECSahar}
S.~Alipour, A.~T. Rezakhani, Extended convexity of quantum fisher information
  in quantum metrology, Physical Review A 91 (2015) 042104.

\bibitem{rossi2014engineering}
M.~A. Rossi, C.~Benedetti, M.~G. Paris, Engineering decoherence for two-qubit
  systems interacting with a classical environment, International Journal of
  Quantum Information 12~(07n08) (2014) 1560003.

\end{thebibliography}

\end{document}